# On the interpretation of the classical GRT tests and cosmological constant in anisotropic geometrodynamics

**Sergey Siparov**

State University of Civil Aviation, 38 Pilotov str., St-Petersburg, 196210, Russia

Research Institute for Hyper Complex Systems in Geometry and Physics, 3 bg 1 Zavodskoy pr., Fryazino, Moscow region, 141190, Russia

E-mail: sergey@siparov.ru

In this paper the basic model for the anisotropic geometrodynamics (AGD) is introduced and used to discuss the classical GRT tests – orbit precession, light bending and gravitational red shift – on the galactic scale. It is shown that the corresponding effects could be used for the interpretation of the observed phenomena. Moreover, the AGD approach provides the interpretations that are free from the need of dark matter and dark energy, and explains the problems with the cosmological constant which is introduced when the problems arise.

1. Introduction

Recently there was regarded [1] the way to solve the astrophysical problems following from observations and contradicting the classical GRT. These problems were listed in [2] as the observational restrictions that must be sufficed in any modification of the gravitation theory: flat rotation curves of spiral galaxies; Tully-Fisher law for the luminosity of such galaxies; particular features in the motion of globular clusters, namely, the classical character of this motion in the direction orthogonal to the galaxy plane (differing from the situation with stars moving in the galactic plane for which the theory modification is needed) and the observed concentration of these clusters in the vicinity of the galaxy center that contradicts the Keplerian expectations for their location. Besides, there is a problem of an essential quantitative excess in the observed gravitational lensing effect in comparison with the GRT prediction. The brief review of the attempts of the needed modification of the theory was given in [1, 3]. The most successful (empirically rooted) theory, which is MOND, manages to fit most of the observations on the base of arbitrary choice of functions but leads to errors in the prediction of other observable phenomena [2].

In [1] there was suggested an approach that generalizes the GRT for the case of the moving distributed sources of gravitation. The basic idea of it is that any theory that combines Newtonian gravity (inverse square law) together with Lorentz invariance in a consistent way must include a field, which is generated by mass currents. Therefore, the geometrical identity

$\frac{\partial F_{ij}}{\partial x^k} + \frac{\partial F_{jk}}{\partial x^i} + \frac{\partial F_{ki}}{\partial x^j} = 0$ in which $F_{jt} = \frac{\partial A_t}{\partial x^j} - \frac{\partial A_j}{\partial x^t}$ is just an anti-symmetric tensor can be used for deriving Maxwell equations both for electromagnetism (EM) and gravitation. This is related to the approach known as gravitoelectromagnetism (GEM) [4] but, obviously, it should not be that straightforward. The GEM theory copies the electromagnetism, while the velocity dependent Lorentz force which is present in EM is produced by an additional, external field. According to the spirit of the GRT, the gravitation force must be not external one but has to enter the metric due to the equivalence principle. Instead of Riemannian isotropic metric, an anisotropic one was introduced which lead to the velocity dependent gravitational forces. The metric suggested in [1] has the form

$$\tilde{g}_{ij}(x, u(x), y) \equiv g_{ij}(x, y) = \gamma_{ij} + \varepsilon_{ij}(x, y) \tag{1.1}$$

where $\gamma_{ij}$ is x-independent metric (here: Minkowski one); $\varepsilon_{ij}(x, y)$ is a small anisotropic perturbation; $y$ belongs to the tangent space, and along some curve (trajectory of the probe particle) $x^i = x^i(s)$, we shall always consider $y^i = \frac{dx^i}{ds}$; finally, $u(x)$ is the vector field corresponding to the motion of sources, and it generates the anisotropy. It seems natural to call this approach the *anisotropic geometrodynamics* (AGD). The geometry endowed with metric of eq. (1.1) type is known as generalized Lagrangean one.

Varying the Lagrangian $L = (\gamma_{hl} + \varepsilon_{hl}(x, y)) y^h y^l$, the Euler-Lagrange equations and the generalized geodesics were obtained similarly to [5]. The generalized geodesics is

$$\frac{dy^i}{ds} + (\Gamma^i_{lk} + \frac{1}{2} \gamma^{it} \frac{\partial^2 \varepsilon_{kl}}{\partial x^j \partial y^t} y^j) y^k y^l = 0 \tag{1.2}$$

where $\Gamma^i_{jk} = \frac{1}{2} \gamma^{ih} (\frac{\partial \varepsilon_{hj}}{\partial x^k} + \frac{\partial \varepsilon_{hk}}{\partial x^j} - \frac{\partial \varepsilon_{jk}}{\partial x^h})$ is y-dependent Christoffel symbol. Omitting the details and discussions given in [1], let us only mention that following the speculations similar to those in [6], the expression for the gravitational force can be transformed into

$$\vec{F}^{(g)} = \frac{mc^2}{2} \left\{ -\nabla_{(x)} \varepsilon_{11} + [\vec{v}, rot_{(x)} \frac{\partial \varepsilon_{11}}{\partial \vec{v}}] + \nabla_{(x)} (\frac{\partial \varepsilon_{11}}{\partial \vec{v}}, \vec{v}) \right\} \tag{1.3}$$

The first term is related to the expression for the usual gravity force, $F^{(g)}_N$, acting on a particle with mass, $m$. For the stationary point source of gravitation with mass $M$, one may choose $\varepsilon_{11}(x) = \frac{r_S}{r}$ where $r_S = \frac{2GM}{c^2}$ corresponds to Schwarzschild radius, and get exactly the Newton formula. If one chooses

$$\Omega(x) = \frac{c^2}{4} rot_{(x)} \frac{\partial \varepsilon_{11}}{\partial \vec{v}}, \tag{1.4}$$

the second term in eq. (1.3) can be recognized as related to a rotating frame Coriolis force $F^{(g)}{}_C = 2m[\vec{v}, \vec{\Omega}]$ which is proportional to velocity $\vec{v}$, of the probe particle and to the proper motion of the gravitation sources. Notice, that the actions produced by the gravitational force component $F^{(g)}{}_C$ on a body could be attraction, repulsion and tangent action depending on the angle between $\vec{v}$ and $\vec{\Omega}$. The component of velocity, $\vec{v}$, which is parallel to $\vec{\Omega}$ is not affected by the second term in eq.(1.3), and this corresponds to one of the features of the globular clusters behavior mentioned above. The third term in eq. (1.3) which is interesting especially in view of the action produced on a moving particle by radial expansion (explosion) or by radial contraction (collapse) of the system of gravitating sources will be discussed in this paper.

Thus, the characteristic features of the AGD approach given in [1] appeared to be the following. The total force acting on the probe particle depends not only on the location of distributed masses but also on their proper motion and on the motion of the particle itself. The gravitational interaction ceased to be simple attraction as before, it depends on the motion of the particle and of the sources and can be attraction, repulsion and transversal action. It goes without saying that all the GRT results remain valid for a planetary system scale on which there is only one essential source of gravitation.

## 2. Basic model of the gravitation source in AGD and its applicability

The basic model of the gravitation source in the GRT is a massive point or a massive spherical body, everything else is corrections. This is why it works well for a planetary system with a star in the center, but fails to explain the flat rotation curves in spiral galaxies. The situation changes in the AGD developed for the distributions of moving sources. For example, a spiral galaxy consists of billions of stars and has the natural preferential direction – the axis of its rotation. It can not be modeled by a point unless we are far enough from it. The basic (and the simplest) model in case of the AGD must be different.

Let a system consist of a central mass and an effective circular mass current, $J^{(m)}$ around it. For brevity, we will call such model a CPC (center plus current). For galaxies like M-104 (Sombrero) or NGC-7742 (Fig.1a,b) with the pronounced ring structure, this model can be used at once. For other galaxies – with emphasized spiral arms – the effective values of contour radius, $R_{eff}$, constant angular velocity, $\Omega_{eff}$, and linear velocity of mass density motion along the contour, $V_{eff} = \Omega_{eff} R_{eff}$, should be introduced. It could be done, for example, in the following way

$$I_{eff} = \sum I_n \equiv M R_{eff}^2 \Rightarrow R_{eff}^2 = \sqrt{\frac{I_{eff}}{M}} \qquad (2.1)$$

where $I_{eff}$ is the moment of inertia of the system with the total mass, $M$. The effective angular velocity, $\Omega_{eff}$, can be defined from $I_{eff}\Omega_{eff} \equiv L_{eff} = \sum L_n$, where $L_n$ is the angular momentum of the component of the system. We get, thus,

$$\Omega_{eff} = \frac{L_{eff}}{I_{eff}} \qquad (2.2)$$

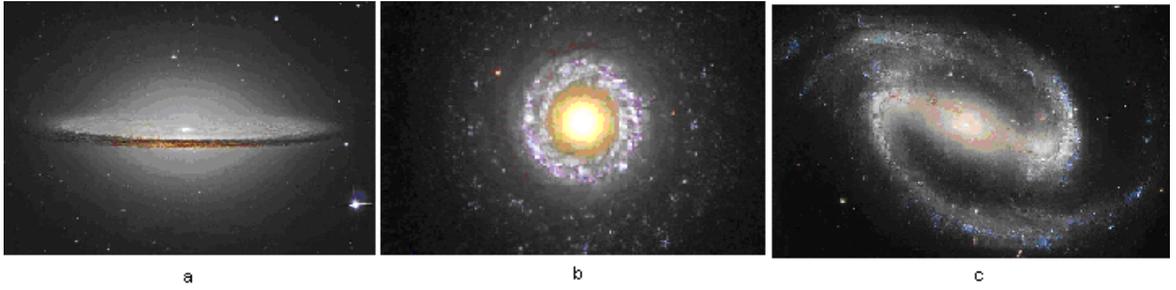

Fig.1. Spiral galaxies: a – M-104, b – NGC-7742, c – NGC-1300

These parameters can be estimated for a chosen galaxy from the astronomical observations.

Due to the mentioned above identity of the origin of Maxwell equations for electrodynamics and for gravitation, such model is quite similar to electromagnetic one consisting of a circular electric current and a charge at the center of it. Thus, if we neglect for a while the third term in eq. (1.3) (the possibility of this will be discussed later), the mathematical results already obtained in electrodynamics can be used in calculations dealing with velocity dependent gravitation.

In [1] this model was used to obtain the flat rotation curves, Tully-Fisher law, to explain the mentioned features of the globular clusters behavior and to discuss the causes of the quantitative discrepancies between the predicted and observed refraction by a gravitational lens. The region of the applicability of this model was estimated and appeared to have a good coincidence with the observed radii of spiral galaxies and with the velocities of stars on their peripheries. The dark matter free explanation of the flat rotation curves makes one doubt if this notion is needed in other cases that one could try to interpret on the base of the AGD.

### 3. Classical GRT tests in the AGD

The CPC model makes it possible to discuss the classical GRT tests – orbit precession, light bending and gravitational red shift – as they appear on the galactic scale. In view of the AGD approach, some visual results can be obtained with the help of numerical calculations. The last is primarily due to the fact that the needed functions (known for the CPC model in electrodynamics) are expressed by elliptical integrals.

*Orbit precession*

In order to describe the motion of a star around a spiral galaxy with a bulge in the center, let us use the CPC model and regard a particle orbiting the system in the plane of the contour. Strictly speaking, a particle in such a system can not perform finite motion and has either to fly away or to fall on the center. Therefore, as can be seen on Fig.2a, there are good reasons to find the globular clusters in the vicinity of the center and not in the periphery where they should spend the majority of their lifetime according to Keplerian elliptical orbits corresponding to Newton gravity. If the number of the particle's rotations characterizing the meta-stable situation when it orbits the CPC is large enough, one can see on Fig.2b what could be called the quasi-precession. This type of star's motion is a galactic scale AGD analogue of the GRT orbit precession. It is also this type of motion that is needed for the explanation of the spiral's arms origination by the density wave theory [7].

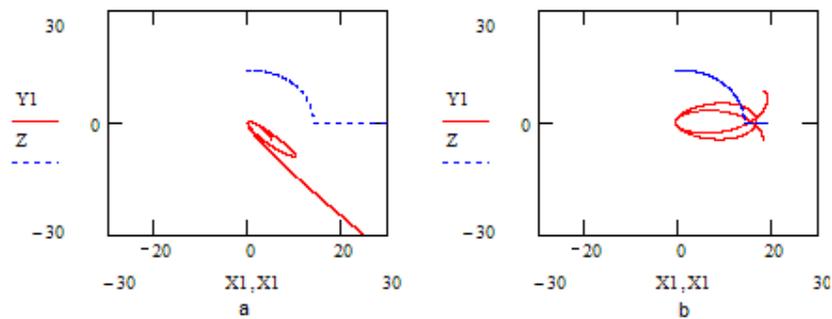

Fig.2. AGD based trajectories mimicking various observable phenomena: a – non-Keplerian behavior of globular clusters; b – quasi-precession

*Light bending*

On Fig.3 there are the results of the numerical modeling of a probe body scattering on a

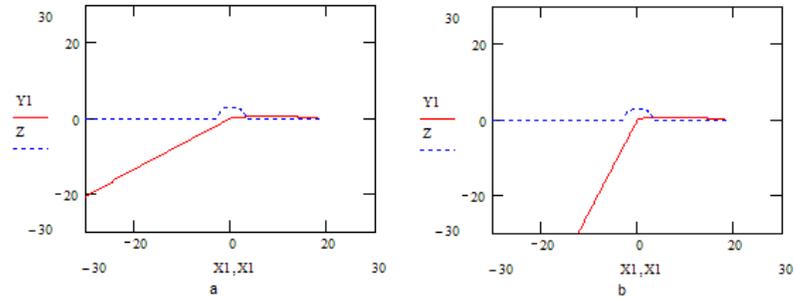

Fig.3. AGD based trajectories for a probe body scattering: a – on the Coulomb center; b – on the CPC

Coulomb center and on a CPC. Since the metric now depends on the proper motion of the source, the light beams will give similar trajectories. This explains why the observable gravitational lenses can give larger refraction than they should according to the GRT in which another basic model is used to model them. Moreover, on Fig.4 there is a probe body trajectory with double bending which appears with the change of the current direction. From the point of view of light refraction, it means that the AGD predicts the existence of the negative gravitational lenses that could diminish the angular size of the objects behind them. Therefore, if such a lens is located between the object and the observer on the Earth, the distance to the object might be considered larger than it really is. If this object is the astronomical standard candle, then this could be the reason of misinterpretation of the last decade supernovae 1a observations [8] as breaking the linear Hubble law. It is these observations that lead to the appearance of the notion of dark energy (of repulsion) that causes the acceleration of the Universe expansion, and to the revival of the cosmological constant in Einstein equations. Now it has to have the opposite sign with regard to that used by Einstein who also needed extra energy (of attraction) to preserve the observable stationary Universe consistent with the GRT. But interpreting these SN1a observations with regard to the possible negative gravitational lenses refraction predicted by the AGD, one has no need to demand this dark energy from the Universe, while all the results of the GRT remain still valid in the region of its applicability.

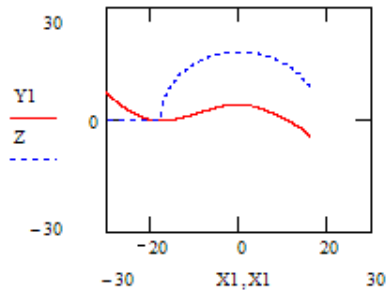

Fig.4. AGD based trajectory with double bending

*Gravitational red shift*

Similarly to the GRT, the gravitation in the AGD also causes the increase of the time interval and consequently, causes the gravitational red shift, but as follows from eq. (1.1) now it also depends on the motion of the source of gravitation. This means that the light coming to the Earth from the far away can be affected by the masses moving tangentially to the line of sight and can acquire the gravitational red shift. This suggestion finds support in the observations of the tangent motions of distant quasars – they take place at amazingly high velocities as it can be seen on Fig.5 taken from [9]. Can this red shift explain the observable red shift discovered by Hubble, is an open question. If its value is comparable to the Hubble red shift which is presumably caused by the Universe expansion, then the very expansion can be doubted. At any rate, this AGD mechanism should be included into the cosmological picture which now becomes not stationary as in Einstein times but also not simply radially expanding as it was considered up to now. As follows from the AGD, the tangential motion and the corresponding gravitation at the periphery of the observed Universe provides a new insight on Mach's principle and on the border of the Universe problem.
.

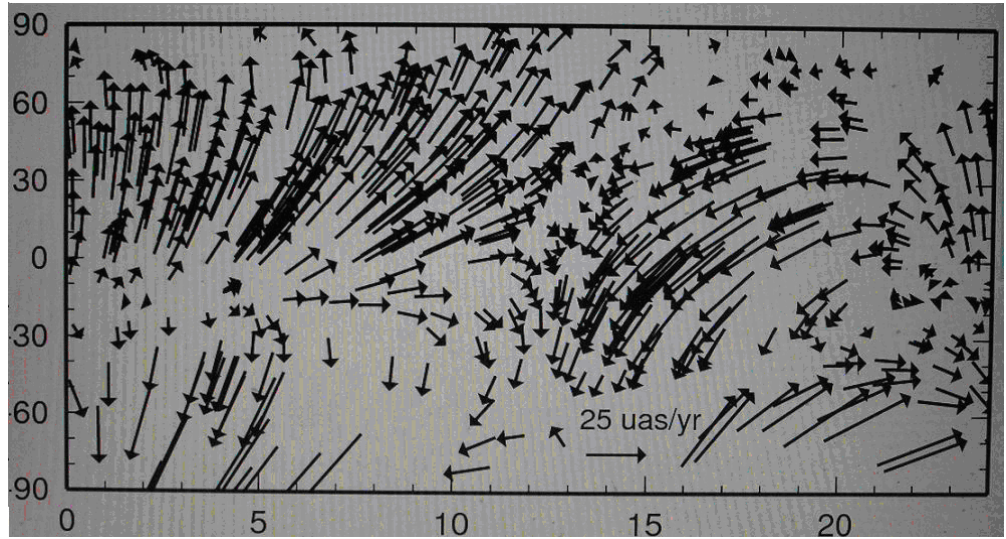

Fig.5. Quasar apparent proper motion [9].

**4. Cosmological constant and the AGD**

Let us investigate the situation with the cosmological constant having this or that sign in order to find out how to understand it in frames of the AGD. First, let us regard a non-inertial reference frame and choose the uniformly rotating one. The force acting on a particle in such a system can be expressed as

$$\vec{F}_{N-I} = m\nabla(\Phi + \frac{u^2}{2}) + m[\vec{v}, rot\vec{u}] \qquad (4.1)$$

where $\Phi$ is the force potential, $v$ is the particle velocity, and $\vec{u} = [\vec{\omega}, \vec{r}]$ is the linear velocity at the distance $r$ from the axis of rotation. The second term in eq. (4.1) is proportional to Coriolis force and the second term in brackets under gradient is called potential centrifugal energy. As $r$ approaches infinity, this energy also becomes infinite which shows the difference between the physical energies (e.g. the energy of gravitational interaction that vanishes at infinity) and formal ones originating from the formalism.

Let us introduce a new designation, $u$, which was already used in eq.(1.1)

$$\vec{u} \equiv \frac{c^2}{2} \frac{\partial \varepsilon_{11}}{\partial \vec{v}} \equiv [\vec{\omega}, \vec{r}] \qquad (4.2)$$

Recollecting that $\varepsilon_{11}$ is the variable part of the metric, one can notice the relation of this definition to the Cartan tensor. On the other hand, it has obvious relation to the corresponding "vector potential" whose physical meaning becomes now more transparent. Now transform the expression eq. (1.3) for the gravitational force into

$$\vec{F}^{(g)} = m\nabla_{(x)}\{-\frac{c^2}{2}\varepsilon_{11} + (\vec{u}, \vec{v})\} + m[\vec{v}, rot_{(x)}\vec{u}] \qquad (4.3)$$

and compare eq. (4.3) with eq. (4.1). For the basic CPC model regarded in Section 2, $u = V_{eff}$ is the linear velocity of the mass current circulating around the central body at the distance $R_{eff}$ from it. Notice, that $\omega = 2\Omega$. We see that in eq. (4.3) the second term under gradient has obtained the features of the physical and not formal energy. It does not become infinite at infinity but can be *finite* or zero there and can have *both* signs depending on the value of the scalar product. Moreover, the meaning of this "infinity" as it appears now in the physical problem that concerns not the vicinity of a single gravitation source but the situation inside the *distribution* of the (moving) gravitation sources requires some clarification.

Let us leave aside for a while the second term in eq. (4.3) and enumerate the possible situations with the probe body interacting with a moving mass distribution.

1. The velocity of the mass current $u$ has no radial component. This means that we deal with the stationary spiral galaxy, and the CPC model is applicable.
    a) The radial component of the probe velocity does not affect the situation.
    b) If the tangent component of the probe velocity gives positive scalar product, an additional repulsive force appears and pushes the probe outwards.
    c) If the tangent component of the probe velocity gives negative scalar product, an additional attraction force appears and pulls the probe inwards.
2. The velocity, $u$, of the mass current has radial component directed outwards with regard to the distribution center (expansion, explosion). This means that besides the consequences b) and c) mentioned in item 1, the following will also take place

- If the radial component of the velocity of the probe is directed outwards, the additional repulsive force appears.
- If the radial component of the velocity of the probe is directed inwards, the additional attraction force appears.

3. The velocity, *u,* of the mass current has radial component directed inwards with regard to the distribution center (contraction, collapse). This means that besides the consequences b) and c) mentioned in item 1, the following will also take place
   - If the radial component of the velocity of the probe is directed outwards, the additional attraction force appears.
   - If the radial component of the velocity of the probe is directed inwards, the additional repulsive force appears.

One can follow the signs of the additional forces appearing in the AGD and compare them to the signs of the cosmological constants that appear in the GRT equations when this or that observation starts to contradict the theory. Notice that the scale of the physical problem discussed here has become essentially larger than before. One could think of the possibility to apply the same approach on the next – galactic clusters – scale. Recollecting that according to the observations, the global structure of the Universe contains voids and walls, one could think again of the preferential directions with regard to these walls and consider similar ideas. Therefore, the distance at which one has to account for the difference between the "formal" infinity and the "physical" one depends on the scale of the problem, while the appearance of the absolute "cosmological constants" characterizing this infinity does not seem natural. Instead (or alongside) with global expansion of the Universe, the AGD suggests more local expansions and contractions depending on the relative motions of the parts of Universe, be it stars, galaxies or galactic clusters.

5. **Discussion**

The main goal of this paper is the discussion of the recently suggested modification of the gravitation theory which is called anisotropic geometrodynamics or AGD and which works at larger scales where the single spherical source model of the GRT starts to fail. Really, all the astrophysical restrictions stemming from observations that must be sufficed by any modification of the gravitation theory are now satisfied. The main idea of the AGD which makes it work is the consistent use of the equivalence principle which actually demands to include all the inertial forces (velocity dependent ones as well) into the metric. This makes it principally different, e.g. from the GEM theory, in which some of them were inside the metric and some were considered external ones like those in the electrodynamics where the additional field is present. Therefore, the metric and the space itself become anisotropic and the suitable geometry is now not Riemannian but the generalized Lagrangean one. But from the physical point of view, it appears that it worth, and all the phenomena known as three classical tests of the GRT appear to be present in the AGD with regard to the interpretations of the observations. The role of the cosmological constant that was every time used to adjust the observations to the theory is discussed. Instead of it, a new term whose sign depends on the physical situation appears. Moreover, it seems possible to do without such notions as dark matter and dark energy, because the AGD suggests dark notions free explanations to the observational data. If the tangential motions of the massive objects at the periphery of the Universe provides the gravitational red shift comparable to Hubble one, new cosmological ideas could be suggested and discussed.

**Acknowledgement**

The work was supported by the RFBR grant No. 07-01-91681-RA_a.